\documentstyle[12pt,epsf]{article}

\newcommand{\gsim}{\;\rlap{\lower 3.5 pt \hbox{$\mathchar \sim$}} \raise 1pt
 \hbox {$>$}\;}
\newcommand{\lsim}{\;\rlap{\lower 3.5 pt \hbox{$\mathchar \sim$}} \raise 1pt
 \hbox {$<$}\;}

\begin{document}
\noindent
\thispagestyle{empty}
\renewcommand{\thefootnote}{\fnsymbol{footnote}}
\begin{flushright}
{\bf TTP99-12}\footnote{The 
  complete postscript file of this
  preprint, including figures, is available via anonymous ftp at
  www-ttp.physik.uni-karlsruhe.de (129.13.102.139) as /ttp99-12/ttp99-12.ps 
  or via www at http://www-ttp.physik.uni-karlsruhe.de/cgi-bin/preprints.}\\
{\bf DESY 99-031}\\
{\bf hep-ph/9903322}\\ 
{\bf March 1999}\\
\end{flushright}

\begin{center}
\begin{Large}
Axial Contributions at the Top Threshold\\
\end{Large}

\vspace{0.5cm}

\begin{large}
J.H.~K\"uhn\\[2mm]
\end{large}
Institut f\"ur Theoretische Teilchenphysik, Universit\"at Karlsruhe, 
D-76128 Karlsruhe, Germany\\  
\vspace{4mm}
\begin{large}
T.~Teubner\\[2mm]
\end{large}
Deutsches Elektronen-Synchrotron DESY, D-22603 Hamburg, Germany\\
\vspace{1.cm}
{\bf Abstract}\\
\vspace{0.5cm}

\noindent
\begin{minipage}{13.0cm}
\begin{small}
We calculate the contributions of the axial current to top quark pair 
production in $e^+ e^-$ annihilation at threshold. The QCD dynamics is taken 
into account by solving the Lippmann-Schwinger equation for the $P$ wave 
production using the QCD potential up to two loops. We demonstrate that the 
dependence of the total and differential cross section on the polarization 
of the $e^+$ and $e^-$ beams allows for an independent extraction of the 
axial current induced cross section.
\end{small}
\end{minipage}

\end{center}

\setcounter{footnote}{0}
\renewcommand{\thefootnote}{\arabic{footnote}}
\vspace{1.cm}

Top quark production at an electron-positron collider \cite{PRep} has been 
demonstrated to be ideally suited for a precise determination of the top
quark mass and for the study of its couplings in production and decay. Due
to its rapid decay large distance nonperturbative QCD effects are
irrelevant for the description of the top quark \cite{K}, and 
the $t\bar t$ system is well described by perturbative QCD \cite{FK}. It
allows to explore the interquark potential at small distances, which is
closely related to the strong coupling constant. One might eventually even 
become sensitive to the $t$-$\bar t$-Higgs coupling through virtual
corrections. In order to constrain this multitude of parameters in an
optimal way and to reduce inevitable theoretical uncertainties, it is 
desirable to measure a large variety of different observables. Originally the 
main emphasis had been put on the total cross 
section \cite{FK, StrasslerPeskin}. The excitation curve 
with its steep rise (the remnant of the 1$S$ toponium resonance) is 
ideally suited for the measurement of the top quark mass $m_t$. The 
correlation between $m_t$ and the strength of the potential ($\alpha_s$) can 
be reduced by comparing data and predictions for the momentum distribution 
of the top quarks \cite{Sumino, JKT, JT, MM}, which reflects essentially 
their Fermi motion in the bound state 
and the smearing of the momentum due to the large decay 
rate $\Gamma_t$, a consequence of the uncertainty principle. All these 
quantities were calculated for the $S$ wave amplitude, which is induced by 
both the electromagnetic current and the vector part of the neutral current 
close to threshold. Expanding in the limit of small velocities 
$\beta = \sqrt{1-4m_t^2/s}$ ($\sqrt{s}$ being the total centre of mass 
energy), the next term is due to $S-P$ wave interference. The subleading 
$P$ wave amplitude originates from the production through the axial vector 
current. 
The interference term is responsible for the anisotropic angular dependence, 
specifically the term linear in $\cos\theta$, and the resulting 
forward-backward asymmetry \cite{Sumino2}. Similarly, an angular dependent 
polarization of top quarks is induced by the $S-P$ wave interference which 
adds to the dominant polarization parallel to the $e^+ e^-$ 
beams~\cite{HJKT}. Rescattering corrections \cite{HJKP, PS}, 
although important for the detailed quantitative analysis, do not alter this
qualitative picture.

Clearly, the next step in this sequence of improvements are corrections of 
order $\beta^2$ which, for interacting quarks close to threshold, 
translate into corrections of order $\alpha_s^2$ and $\beta\alpha_s$.
For the vector current this has been recently persued by different groups, 
which have demonstrated the importance of these next-to-next-to leading 
order corrections \cite{HT3, MYel, Yakovlev}. However, in the same 
order $\beta^2$ (or 
$\alpha_s^2$) also axial vector induced contributions must be incorporated. 
They affect both the excitation curve and the momentum distribution. 
Close to threshold these axial contributions are suppressed relative to 
the dominant $S$ waves by two powers of $\beta$ whence a treatment of 
the leading terms is sufficient for the present purpose. These 
axial contributions are mediated by the virtual $Z$ boson only. Therefore 
their dependence on the beam polarization differs from the one of the vector 
current induced rate. This, in turn, allows for the separation of the two 
independent contributions to the total and differential cross section.
With the axial contribution representing an independent observable, this 
separation is possible independent of potential uncertainties in the NNLO 
calculation of the dominant piece. However, in view of the $\beta^2$ 
suppression of the axial rate and the relatively small couplings of the 
neutral current, large luminosities and a high degree of polarization are 
required to make a clean extraction of the axial part possible. These 
features are unique for linear colliders, as 
proposed e.g. in \cite{PRep, LinearCollider}. However, even without this 
possibility, it will be important to 
control the impact of this contribution on the extraction of the top quark 
mass and the interquark potential. Let us also stress that the axial rate, 
although closely related to the $S-P$ wave interference piece, is an
independent observable. Rescattering corrections, which are present in the 
angular distribution and in the top polarization, are calculated to 
${\cal O}(\alpha_s)$ \cite{HJKP, PS} but shown to be unimportant as long 
as the total cross section is concerned \cite{FKM, MY}. In addition, 
rescattering corrections do not affect the separation between axial and 
vector contributions.

$P$ wave threshold production of massive quarks in $\gamma \gamma$ collisions 
has been analysed for the case of a pure Coulomb potential in 
Ref.~\cite{FKK} and much of the general considerations can be taken over 
to the present case.\footnote{See also Refs.~\cite{FK2} and \cite{BFK} for 
related discussions of $P$ wave production of quarks and squarks in 
$e^+ e^-$ collisions near threshold.}
This refers in particular to the treatment of the 
linearly divergent integrals over the momentum distribution and the order 
of magnitude estimates. However, for definite predictions the QCD potential 
with its logarithmically varying coupling strength has to be employed. The 
relative size of the electromagnetic and weak couplings is important for the 
phenomenological analysis, as well as the dependence on the beam polarization. 

The momentum distribution of the top quark including the influence of beam 
polarization can be written in the form
\begin{eqnarray}
\frac{{\rm d}\sigma}{{\rm d}p} & = & \frac{3 \alpha^2 \Gamma_t}{m_t^4}
 \, \left( 1 - P_+ P_- \right) \left[ \left( a_1 + \chi a_2 \right) 
 \left( 1 - \frac{16}{3}\frac{\alpha_s}{\pi}\right) {\cal D}_{S-S}(p, E)
\right.
\nonumber\\
 & & \qquad\qquad\qquad\qquad + 
\left.
\left( a_5 + \chi a_6 \right) 
 \left( 1 - \frac{8}{3}\frac{\alpha_s}{\pi}\right) {\cal D}_{P-P}(p, E)
\right] \,,
\label{eqdsigmadp}
\end{eqnarray}
where the correction factors from hard gluon exchange, 
$\left(1-16\alpha_s/3\pi\right)$ and $\left(1-8\alpha_s/3\pi\right)$, are 
taken from \cite{Barbierietal, KZ}. 
$P_+$ and $P_-$ denote the polarization of the positron and electron beams, 
respectively, and $\chi$ is defined as
\begin{equation}
\chi = {P_+ - P_- \over 1 - P_+ P_-}\,.
\label{defchi}
\end{equation}
The coefficients $a_i$ read
\begin{eqnarray}
a_1 = \left( q_e q_t + v_e v_t d\right) ^2 + \left( a_e v_t d\right) ^2 \,, 
& \qquad &
a_2 = 2 a_e v_t d \left( q_e q_t  + v_e v_t d\right) \,,\nonumber\\[1mm]
a_5 = \left( a_t d\right) ^2 \left( v_e^2 + a_e^2\right) \,, 
\qquad\qquad & \qquad &
a_6 = 2 v_e a_e \left( a_t d\right) ^2\,,
\label{eqcouplingsa}
\end{eqnarray}
with 
\begin{equation}
d = {1\over 16 \sin^2\theta_W\cos^2\theta_W}\,{s\over s - M_Z^2}
\end{equation}
and the electromagnetic and weak charges
\begin{eqnarray}
q_e = -1 \,,\quad & v_e = -1 + 4 \sin^2\theta_W\,,\quad 
& a_e = -1\,,\nonumber\\[1mm]
q_t = 2/3 \,,\quad & v_t = 1 - 8/3\, \sin^2\theta_W\,,\quad
& a_t = 1\,.
\label{eqcharges}
\end{eqnarray}
The dynamics of the strong interaction is contained in the functions
\begin{eqnarray}
{\cal D}_{S-S}(p, E) = p^2 \left| G(p, E) \right|^2
\qquad {\rm and} \qquad 
{\cal D}_{P-P}(p, E) = p^2 \left| \frac{p}{m_t}\,F(p, E) \right|^2 \,,
\label{eqDssDpp}
\end{eqnarray}
where $E = \sqrt{s} - 2 m_t$ is the energy relative to the nominal threshold. 
The $S$ and $P$ wave Green functions $G(p, E)$ and $F(p, E)$ fulfill 
the Lippmann-Schwinger equations
\begin{eqnarray}
G(p,E) &=&
G_0(p,E) +
G_0(p,E)
\int {{\rm d}^3q\over(2\pi)^3}
\,\tilde V\left(\,|\,\vec{p}-\vec{q}\,|\,\right)
G(q,E)\,,
\label{eq:LS1}\\[1mm]
F(p,E) &=&
G_0(p,E) +
G_0(p,E)
\int {{\rm d}^3q\over(2\pi)^3}
{{\vec p}\cdot{\vec q}\over p^2}
\,\tilde V\left(\,|\,\vec{p}-\vec{q}\,|\,\right)
F(q,E)
\label{eq:LS2}
\end{eqnarray}
where $p=|\,\vec p\,|$ is the momentum
of the top quark in $t\bar t$ rest frame,
$G_0(p,E) = \left(\, E- {p^2/ m_t}+ {\rm i}\Gamma_t\,\right)^{-1}$ is the 
free Green function, and $\Gamma_t$ denotes the top quark width. 
For the QCD potential in momentum space, $\tilde V$, we adopt the two loop 
result \cite{PeterSchroeder} with the long distance regularization as 
described in \cite{JKPST}. Eqs.~(\ref{eq:LS1}, \ref{eq:LS2}) are then solved 
numerically as described in \cite{JKT, rhdipl}. For all the results 
discussed below we use the parameters $m_t = 175$ GeV, $\Gamma_t = 1.43$ 
GeV and $\alpha_s(M_Z^2) = 0.118$. 

For large momenta both $G(p, E)$ and $F(p, E)$ approach the free Green 
function $G_0$. It is thus evident that the integral over the momentum 
distribution diverges linearly for the $P$ wave. This is, however, an 
artefact of the 
nonrelativistic approximation. The problem could be cured, for example, by 
introducing in this region the relativistic (free) Green function and phase 
space and by treating the interaction as a (small) perturbation. However, 
in practice, a cutoff will be provided by the experimental analysis. The 
invariant mass of the $W$ plus $b$ jet in events with large $p$ ($W b$) and 
small $E = \sqrt{s} - 2 m_t$ will necessarily be strongly shifted away from 
$m_t$ towards smaller values. Such 
events will either not be included in the $t\bar t$ sample or, in any case, 
will require special treatment. Hence, wherever total cross sections are 
presented, a cutoff $p_{\rm max}$ of order $m_t$ will be introduced which is 
easily included also in the experimental analysis.

\begin{figure}[htb]
\vspace{-1cm}
\begin{center}
\leavevmode
\epsfxsize=11.0cm
\epsffile[100 60 460 740]{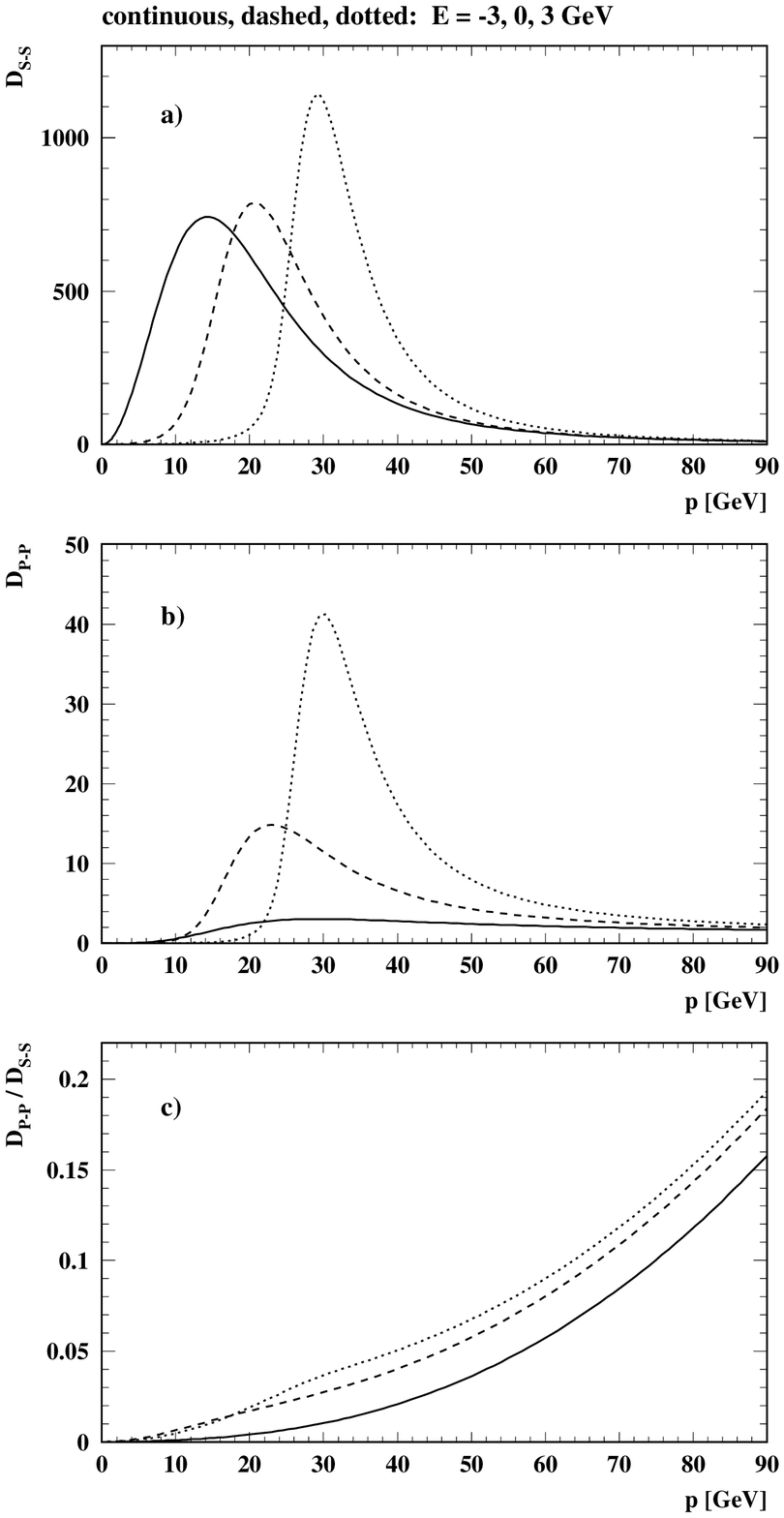}
\end{center}
\vspace{-1cm}
\caption[]{\label{figDssDpp} Results for the basic elements in 
Eq.~(\ref{eqdsigmadp}): a) ${\cal D}_{S-S}(p, E)$, b) ${\cal D}_{P-P}(p, E)$ 
and c) the ratio ${\cal D}_{P-P}/{\cal D}_{S-S}$ for the three energies 
$E = -3$~GeV (continuous curves), $E = 0$ (dashed lines) and $E = 3$~GeV 
(dotted) as a function of the top quark momentum $p$.}
\end{figure}
The relative magnitude of the $P$ wave result is best visualized by 
considering the basic elements ${\cal D}_{S-S}$ and ${\cal D}_{P-P}$ which 
enter Eq.~(\ref{eqdsigmadp}). In Fig.~\ref{figDssDpp} we show these 
distributions for three energies, $E = -3$, $0$ and $3$~GeV. These energies 
roughly correspond to the location of the 1$S$ peak, the nominal threshold 
and the onset of the continuum. For the $S$ wave (Fig.~\ref{figDssDpp}a) we 
observe a fairly wide distribution at $E = -3$~GeV, a consequence of the 
momentum spread of the constituents in the 1$S$ bound state. With increasing 
energy the interaction becomes less important, the width of the distribution 
decreases and approaches the free result $\Gamma_t\sqrt{m_t/E}$. For the $P$ 
wave (Fig.~\ref{figDssDpp}b) the contribution is tiny at $E = -3$~GeV and 
develops a peak only gradually with increasing energy. The ratio 
${\cal D}_{S-S}/{\cal D}_{P-P}$ is shown in Fig.~\ref{figDssDpp}c. For 
energies well above threshold its behaviour is essentially given by the factor 
$p^2/m_t^2$, since both $F$ and $G$ are approximated by the free Green 
function $G_0$. However, for energies relatively close to threshold the 
strong interaction distorts the free wave functions which leads to a 
deviation from the pure $p^2/m_t^2$ behaviour.
\begin{figure}[htb]
\vspace{-0.8cm}
\begin{center}
\leavevmode
\epsfxsize=12.0cm
\epsffile[100 100 460 700]{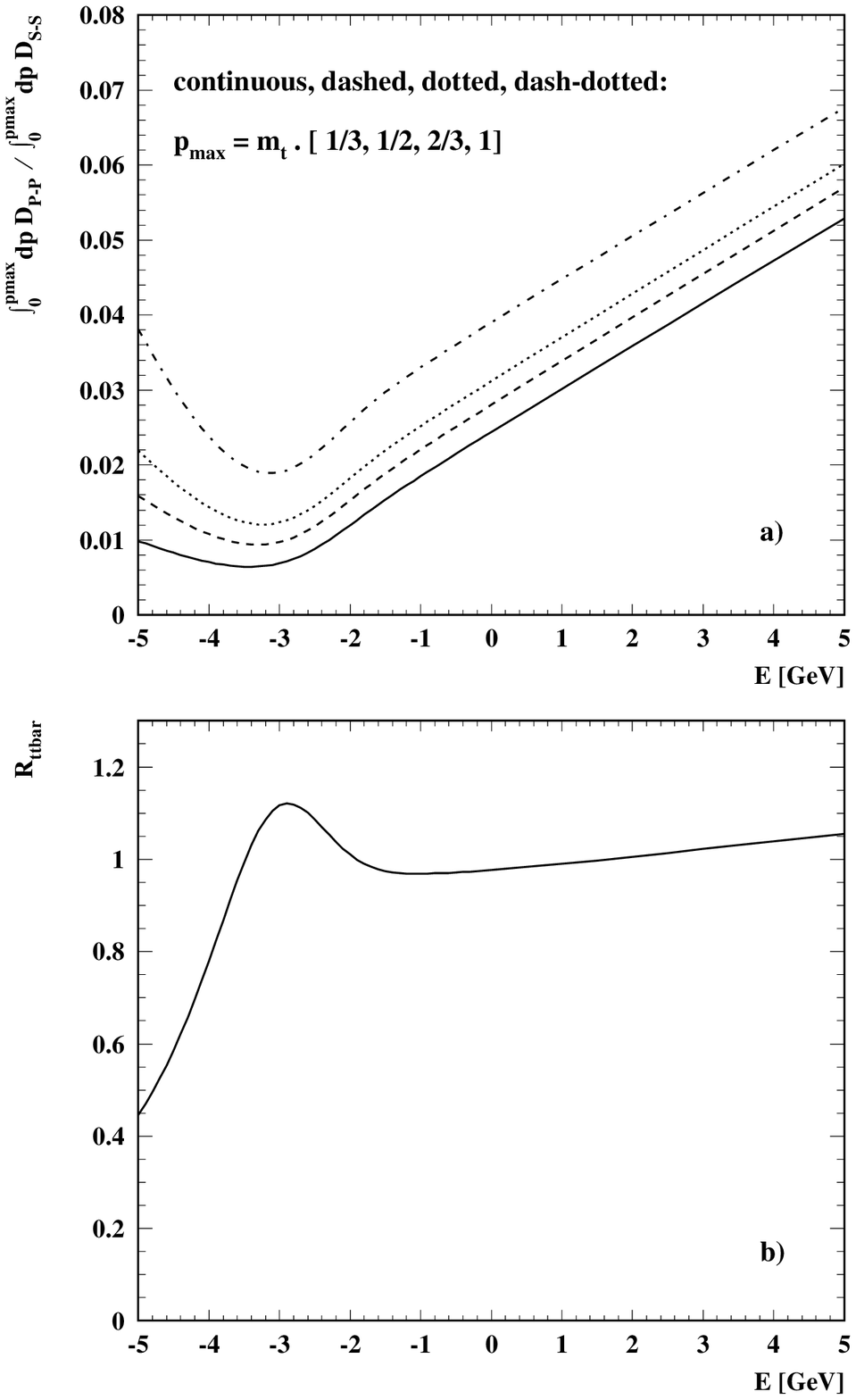}
\end{center}
\vspace{-1.4cm}
\caption[]{\label{figintD}\mbox{}\\
a) Ratio of the integrated distributions: 
$\int_0^{p_{\rm max}}{\rm d}p\,{\cal D}_{P-P}(p, E)\,/\,
 \int_0^{p_{\rm max}}{\rm d}p\,{\cal D}_{S-S}(p, E)$ as a function of 
the energy $E = \sqrt{s} - 2 m_t$ for four different values 
of the cutoff: continuous, dashed, dotted, and dash-dotted lines correspond to
$p_{\rm max} = m_t \cdot \left[\frac{1}{3},\ \frac{1}{2},
\ \frac{2}{3},\ 1\right]$, respectively. 
b) The normalized total cross section $R_{t\bar t}$ as defined in 
Eq.~(\ref{eqdefR}) as a function of $E$.}
\end{figure}
The ratio of the integrated $S$ and $P$ wave distributions as functions of 
$E$ is shown in Fig.~\ref{figintD}a. The different curves give the results 
for different values of the momentum cutoff $p_{\rm max}$ which is applied 
both in numerator and denominator. For a realistic 
analysis $p_{\rm max} = m_t/3$ or $m_t/2$ should be used at most. For free 
and stable quarks the ratio is given by $p^2(E)/m_t^2 \approx E/m_t$. 
Fig.~\ref{figintD}a shows that close to threshold the momentum spread from 
the QCD bound state dynamics leads to a significant modification of the 
$E/m_t$ behaviour and increases the $P$ wave contribution.
The minimum of the ratio $\int_0^{p_{\rm max}}{\rm d}p\,{\cal D}_{P-P}\,/\,
\int_0^{p_{\rm max}}{\rm d}p\,{\cal D}_{S-S}$ is observed roughly at the 
location of the remnant of the 1$S$ peak of the $R_{t\bar t}$ ratio 
(Fig.~\ref{figintD}b)
\begin{equation}
R_{t\bar t} \equiv \frac{\sigma(e^+ e^- \to \gamma^* \to t\bar t\,)}
{\sigma(e^+ e^- \to \gamma^* \to \mu^+ \mu^-)} = \frac{4 \Gamma_t}{\pi m_t^2}\,
\int_0^{p_{\rm max}} {\rm d}p \, {\cal D}_{S-S}\,.
\label{eqdefR}
\end{equation}
\begin{figure}[htb]
\begin{center}
\leavevmode
\epsfxsize=8.0cm
\epsffile[130 70 430 750]{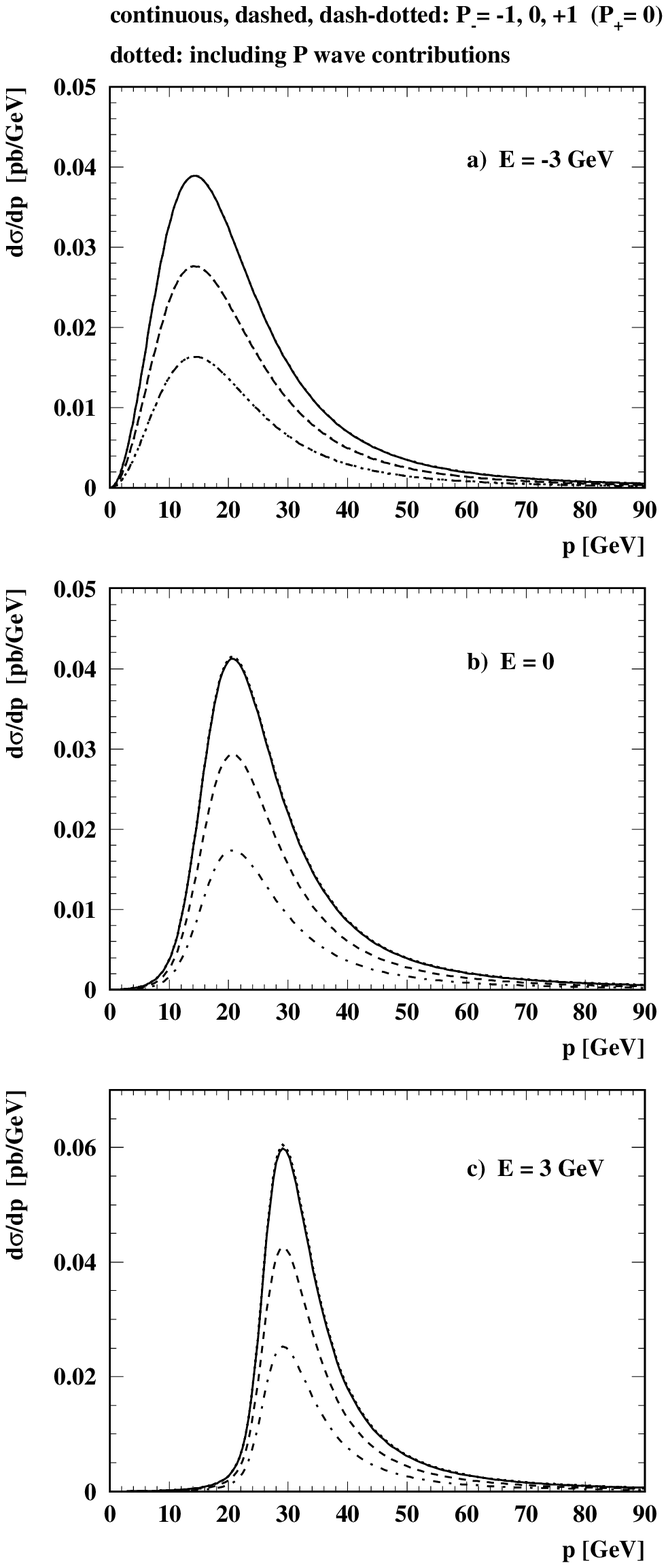}
\leavevmode
\epsfxsize=8.0cm
\epsffile[130 70 430 750]{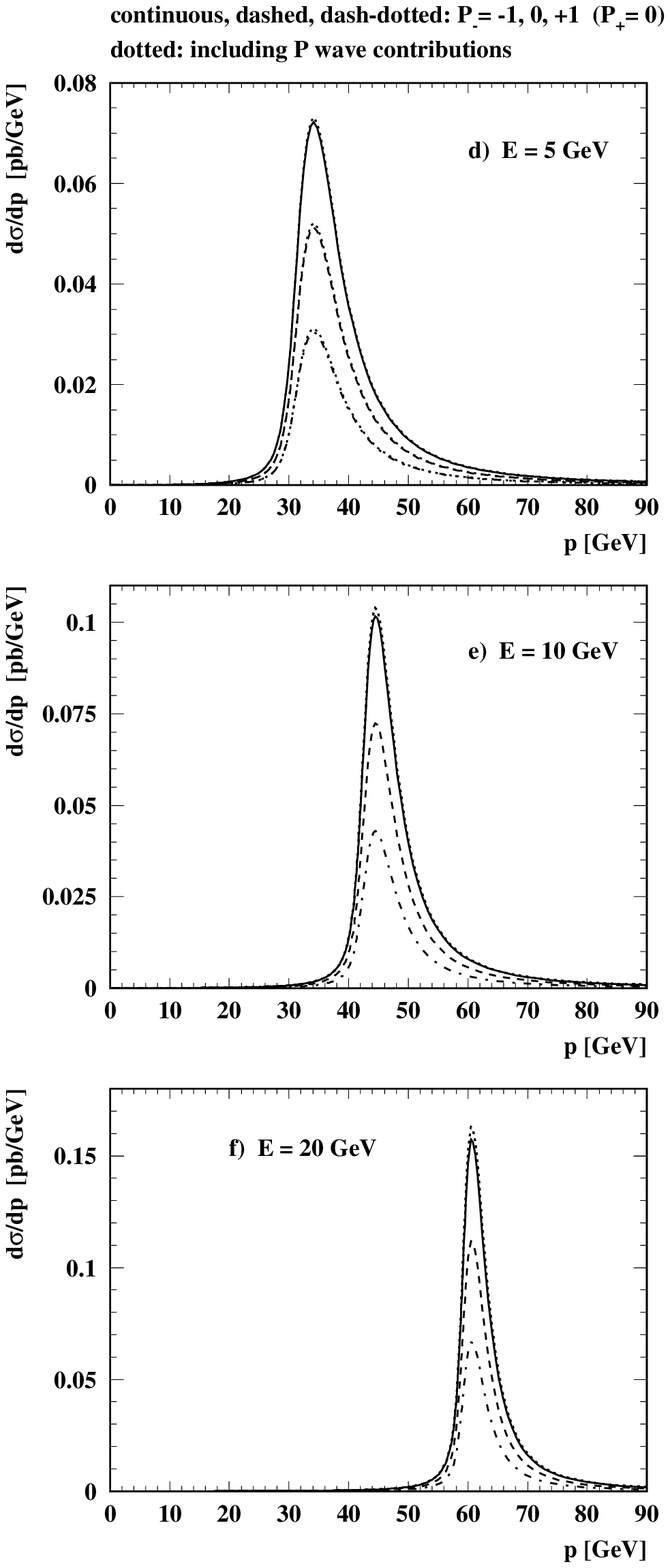}
\end{center}
\caption[]{\label{figdiff} Differential cross section 
${\rm d}\sigma(e^+ e^- \to t\bar t\,)/{\rm d}p$ as defined in 
Eq.~(\ref{eqdsigmadp}) as 
a function of $p$ for six different energies, $E = -3, 0, 3, 5, 10, 20$ GeV, 
as indicated in the plots a)$\ldots$ f). The continuous, dashed and 
dash-dotted lines show the pure $S$ wave result for the three different 
choices of the $e^-$ polarization $P_- = -1$, $0$ and $1$, 
respectively. ($P_+ = 0$.) 
The dotted lines show the full result including the $P$ wave contributions.}
\end{figure}
\begin{figure}[htb]
\begin{center}
\leavevmode
\epsfxsize=8.0cm
\epsffile[130 70 430 730]{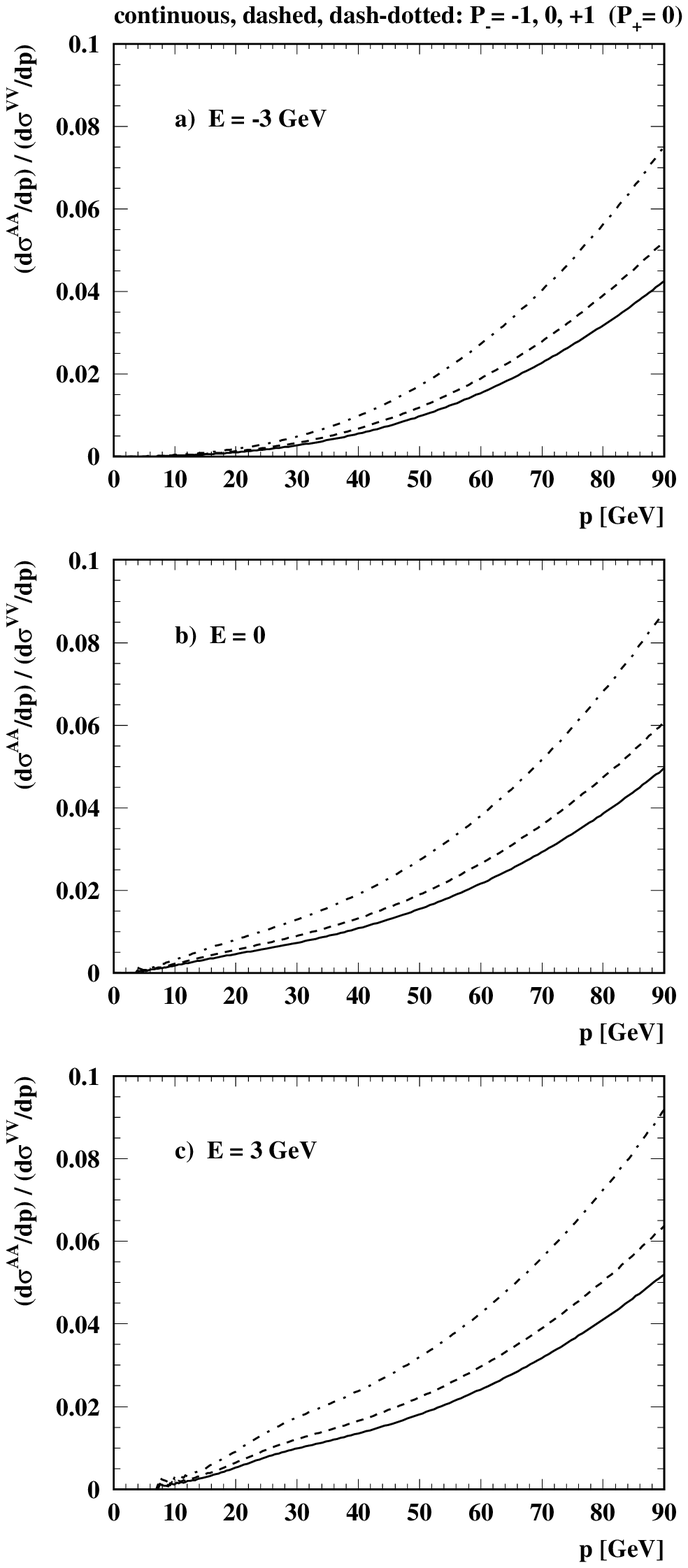}
\leavevmode
\epsfxsize=8.0cm
\epsffile[130 70 430 730]{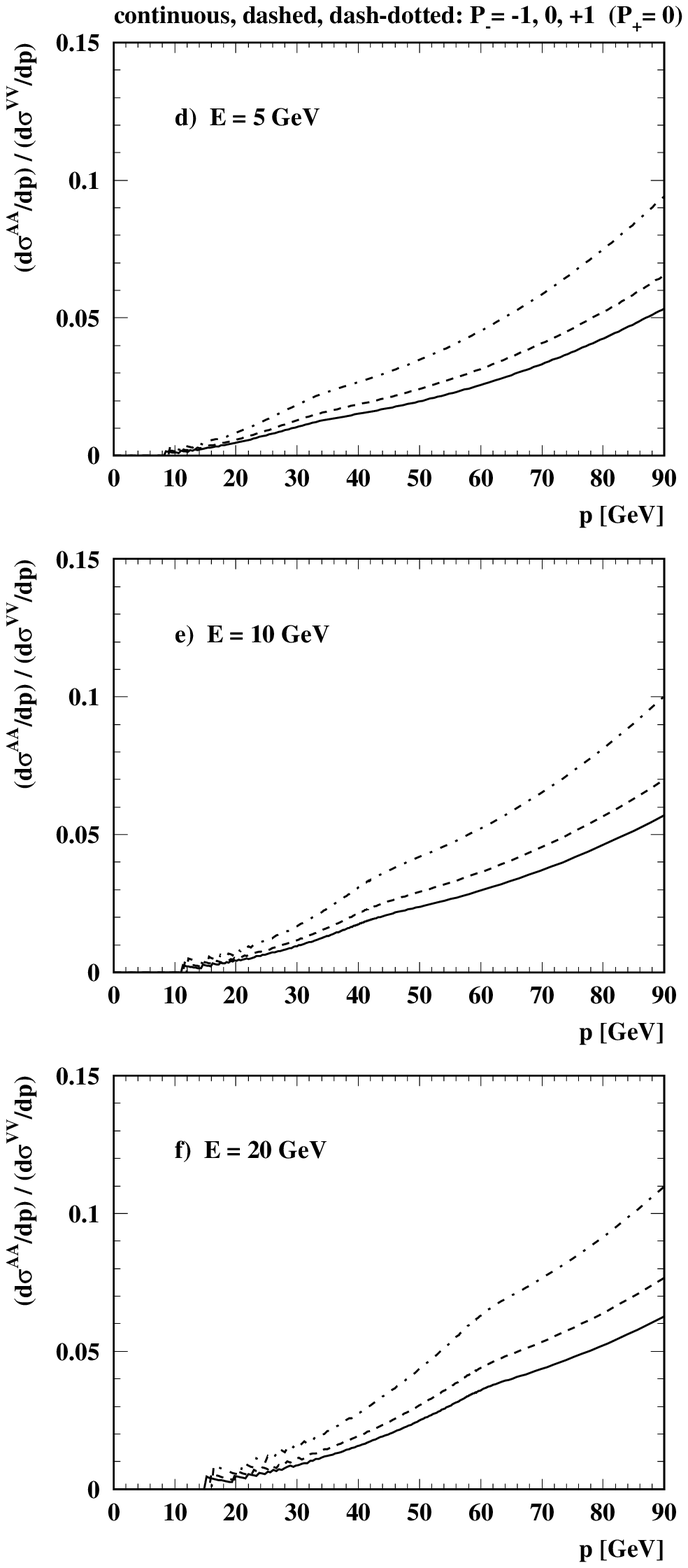}
\end{center}
\caption[]{\label{figdiffr} Relative size of the axial contribution 
${\rm d}\sigma^{\rm AA}/{\rm d}p$ compared to the vector contribution 
${\rm d}\sigma^{\rm VV}/{\rm d}p$ to the differential cross 
section as a function 
of $p$ for six different values of the energy $E$ [plots a)$\ldots$ f)]. The 
continuous, dashed and dash-dotted lines correspond to $P_- = -1$, $0$ and 
$1$, respectively. ($P_+ = 0$.)}
\end{figure}
\begin{figure}[htb]
\begin{center}
\leavevmode
\epsfxsize=11.0cm
\epsffile[100 120 460 700]{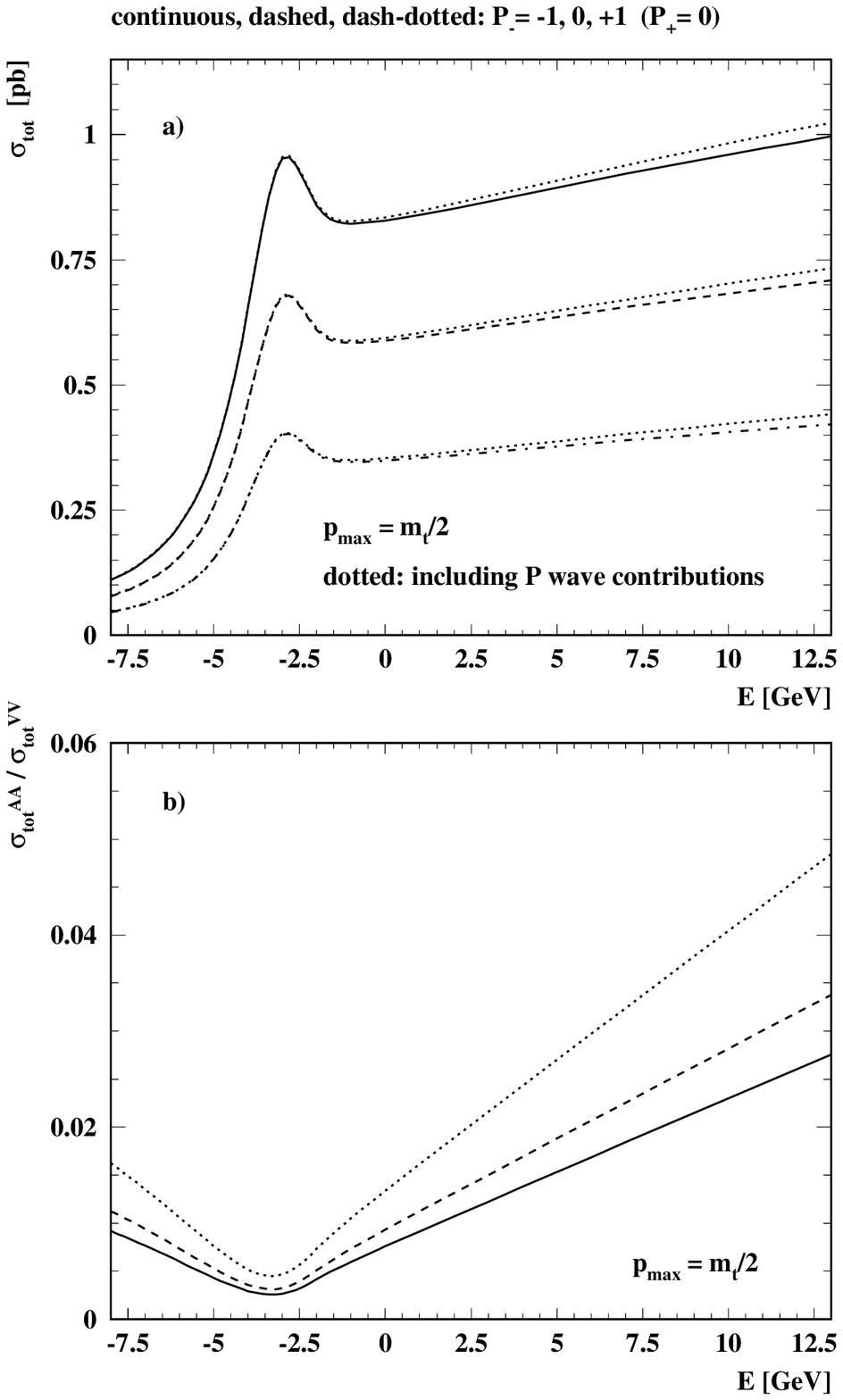}
\end{center}
\vspace{-1cm}
\caption[]{\label{figtotalxsec} a) The total cross 
section $\sigma(e^+ e^- \to t\bar t\,)$ 
as a function of $E$ for three different choices of the $e^-$ polarization: 
the continuous, dashed and dash-dotted lines correspond to $P_- = -1$, $0$ 
and $1$, respectively, where only $S$ wave production is taken into account. 
The dotted lines show the corresponding total cross sections including 
the $P$ wave contributions. b) Ratio of the $P$ to the $S$ wave contribution 
$\sigma_{\rm tot}^{\rm AA}/\sigma_{\rm tot}^{\rm VV}$ for the 
three different $e^-$ polarizations.}
\end{figure}
With these ingredients it is now straightforward to predict the differential 
distribution for the three characteristic polarizations $P_- = -1$, 0, +1 and 
$P_+ = 0$. The cross sections are drastically different for the three choices, 
see Fig.~\ref{figdiff}, reflecting the large left--right asymmetry 
$A_{\rm LR} = a_2/a_1 \approx 0.4$ of the $S$ wave 
contribution~\cite{K,GKKS}. Including the small $P$ wave term (dotted 
curves) leads to marginal changes only, which are barely visible in 
Fig.~\ref{figdiff} even for the highest energies. The relative size of the 
axial contribution is better visible in Fig.~\ref{figdiffr} where the ratio 
between the axial and the vector contribution is plotted as a function of the 
momentum $p$. The shapes and the magnitude are fairly similar for the 
different energies. This is a consequence of the fact, that the ratio 
${\cal D}_{P-P}(p, E)/{\cal D}_{S-S}(p, E)$ is relatively insensitive to the 
energy. In fact, in the absence of interaction this ratio is just given by 
$p^2/m_t^2$, independent of $E$. In contrast, the location of the maximun 
of the distribution itself varies with $E$, and this is mainly responsible
for the increase of the integrated 
$P$ wave cross section. The integrated cross section with and without the $P$ 
wave contribution is shown in Fig.~\ref{figtotalxsec}a, where for the cutoff 
$p_{\rm max} = m_t/2$ is adopted. The ratio between axial and vector 
contributions, both integrated up to $m_t/2$ is shown in 
Fig.~\ref{figtotalxsec}b. 
The shape of these curves reflects the shape of the ratio 
$\,\int {\rm d}p \, {\cal D}_{P-P}\,/\,\int {\rm d}p \,{\cal D}_{S-S}\,$ 
displayed already in 
Fig.~\ref{figintD}a. The normalization depends on the polarization. This 
demonstrates that experiments with polarized beams are able to 
extract $\sigma_{\rm tot}^{\rm AA}$ separately, provided that a 
statistical and systematic precision at the 
percent level can be reached. In any case, if a theoretical prediction of 
shape and normalization of $\,{\rm d}\sigma/{\rm d}p\,$ at a precision of one 
or two percent is needed the $P$ wave contribution to the cross section 
should be \pagebreak[4] included.

\vspace{1cm}
\noindent
{\bf Acknowledgments}\\[1mm]
We thank Robert Harlander for providing us with the implementation of
polarization and $P$ wave contributions in TOPPIK and for useful discussions. 
Work partly supported by BMBF Contract 056 KA 93 P6 at the University of 
Karlsruhe.

\def\prep#1#2#3{{\it Phys.~Rep.~}{\bf #1} (#2) #3}
\def\app#1#2#3{{\it Acta~Phys.~Polon.~}{\bf B #1} (#2) #3}
\def\apa#1#2#3{{\it Acta Physica Austriaca~}{\bf#1} (#2) #3}
\def\npb#1#2#3{{\it Nucl.~Phys.~}{\bf B #1} (#2) #3}
\def\plb#1#2#3{{\it Phys.~Lett.~}{\bf B #1} (#2) #3}
\def\prd#1#2#3{{\it Phys.~Rev.~}{\bf D #1} (#2) #3}
\def\pR#1#2#3{{\it Phys.~Rev.~}{\bf #1} (#2) #3}
\def\prl#1#2#3{{\it Phys.~Rev.~Lett.~}{\bf #1} (#2) #3}
\def\sovnp#1#2#3{{\it Sov.~J.~Nucl.~Phys.~}{\bf #1} (#2) #3}
\def\yadfiz#1#2#3{{\it Yad.~Fiz.~}{\bf #1} (#2) #3}
\def\jetp#1#2#3{{\it JETP~Lett.~}{\bf #1} (#2) #3}
\def\zpc#1#2#3{{\it Z.~Phys.~}{\bf C #1} (#2) #3}

\end{document}